\documentclass[12pt,amstex]{article}
\pdfoutput=1 
\usepackage{graphicx}
\usepackage{rotating}
\usepackage{color}
\usepackage{amssymb,amsmath}
\def\lesssim{\mathrel{\hbox{\rlap{\hbox{\lower5pt\hbox{$\sim$}}}\hbox{$<$}}}}
\def\gtrsim{\mathrel{\hbox{\rlap{\hbox{\lower5pt\hbox{$\sim$}}}\hbox{$>$}}}}

\def\ntrli{\chi^0_i}

\def\chjp{\chi_j^+}

\newcommand{\ntrl}[1]{\chi^0_#1}
\newcommand{\chpm}[1]{\chi^\pm_#1}
\newcommand{\chp}[1]{\chi^+_#1}
\newcommand{\chm}[1]{\chi^-_#1}
\newcommand{\ch}[1]{\chi^{+/-}_#1}
\def\squark{\tilde{q}}

\def\sbotl{\tilde{b}_L}
\def\sbotr{\tilde{b}_R}
\def\stopl{\tilde{t}_L}
\def\stopr{\tilde{t}_R}
\newcommand{\sbot}[1]{\tilde{b}_#1}
\newcommand{\sstop}[1]{\tilde{t}_#1}
\def\squarkc{\tilde{q}^*}

\newcommand{\sbotc}[1]{\tilde{b}_#1^*}
\newcommand{\sstopc}[1]{\tilde{t}_#1^*}
\def\gluino{\tilde{g}}

\def\qbar{\bar{q}}       %
\def\bbar{\bar{b}}       %
\def\tbar{\bar{t}}       %

\def\mtop{m_t}
\def\mbot{m_b}
\newcommand{\mntrl}[1]{m_{\chi^0_#1}}      %

\def\msquark{m_{\tilde{q}}}

\def\msbotl{m_{\tilde{b}_L}}
\def\msbotr{m_{\tilde{b}_R}}
\def\mstopl{m_{\tilde{t}_L}}
\def\mstopr{m_{\tilde{t}_R}}
\def\msq3l{m_{\widetilde{Q}_{3_{L}}}}
\newcommand{\msbot}[1]{m_{\tilde{b}_#1}}
\newcommand{\mstop}[1]{m_{\tilde{t}_#1}}
\def\mgluino{m_{\tilde{g}}}

\def\thetastop{\theta_{\tilde{t}}}
\def\thetasbot{\theta_{\tilde{b}}}

\def\etmiss{\not\!\!{E_T}}

\def\beq{\begin{equation}}   %
\def\eeq{\end{equation}}   %
\begin{document}
\begin{flushright}
   {\bf HRI-P-11-10-002\\ HRI-RECAPP-11-008} 
\end{flushright}
\vskip 30pt
\begin{center}
{\large \bf
Entangled System of Squarks from the Third Generation \\
at the Large Hadron Collider}
%
%
%
\vskip 20pt
AseshKrishna Datta\footnote{asesh@hri.res.in} and 
Saurabh Niyogi\footnote{sourabh@hri.res.in}
\vskip 5pt
{\emph{Regional Centre for Accelerator-based Particle Physics
(RECAPP) \\
Harish-Chandra Research Institute \\
Chhatnag Road, Jhunsi, Allahabad, India 211019} }
\end{center}
\vskip 25pt
\abstract
{In the Minimal Supersymmetric extension of the Standard Model (MSSM) 
squarks from the third generation, i.e., the bottom and the top
squarks, have a common set of parameters that determine their masses
and couplings. These are in the form of the common soft mass for the 
left handed squark from the third generation ($m_{\widetilde{Q}_{_{3L}}}$),
the supersymmetry conserving Higgsino mass parameter $\mu$ and
$\tan\beta$, the ratio of vacuum expectation values of the two Higgs 
doublets. This leads to an interesting possibility that the systems 
involving the bottom and the top squarks might get correlated in a 
non-trivial way even in an unconstrained setup. In this work,
phenomenology of the bottom and the top squarks is studied at the 
Large Hadron Collider which exploits such a possibility with a particular
emphasis on bottom squark decaying to top squark and $W$ boson. 
Possibility of reconstructing multiple $W$ bosons in the final state is
identified to be the key in probing the squark mixing angles.
Further entanglement of the electroweak gaugino-higgsino sector is also
not impossible in scenarios based on phenomenological MSSM 
while for highly constrained scenarios like minimal supergravity/constrained 
MSSM such entanglements are trivially forced upon.
%
\newpage
\section{Introduction}
The Large Hadron Collider (LHC), by entering into the data-taking phase,
has ushered a new era in high energy physics phenomenology. As of now,
each of ATLAS and CMS experiments has collected more than 5.5 fb$^{-1}$ of 
data. Data analysed so far, either published or preliminary in nature and
ranging from 35 pb$^{-1}$ to about 2 fb$^{-1}$, are yet to point out any 
clear evidence of physics beyond the Standard Model (SM) of particle 
physics. Thus, the results presented till date are mostly in terms of 
setting exclusion limits on the otherwise viable regions of the parameter 
space of different plausible scenarios beyond the SM.  For Supersymmetry 
(SUSY) as one such viable frameworks to go beyond the SM, the recent LHC 
analyses 
\cite{atlas-pub-1,atlas-pub-2,atlas-pub-3,atlas-pub-4,
atlas-pub-5,atlas-pub-6,atlas-pub-b1,atlas-prel-b1,atlas-prel-b2,
cms-pub-1,cms-pub-2,cms-pub-3,cms-pub-4,cms-pub-b1,
cms-prel-1,cms-prel-2,cms-prel-3,cms-prel-4,
talk-lepton-photon},
have aggressively extended the region already excluded by Tevatron.

While these are definite and quick improvements over what we knew before 
the LHC data was available, the actual implications of them, at this 
juncture, should be put in perspective. The single most important thing 
the LHC experiments (the ATLAS and the CMS in particular, which are in 
context of the present work) opened up before us is that how powerful and 
efficient these are to reach out to new regimes in the energy scales. 
On the other hand, it is appreciated for quite sometime now that a 
particular SUSY scenario like the minimal Supergravity (mSUGRA) (or, for 
that matter, the constrained version of the Minimal Supersymmetric Standard 
Model (the so-called CMSSM)) in which Tevatron carried out most of its 
analyses and LHC might be sticking to for some more time to come, is at 
best an extremely special but nonetheless, a useful benchmark for any first 
study. So, time is just right to move on to explore the chances of how much 
or what aspects of an (essentially) unconstrained SUSY scenarios 
could, in principle, be put in the context of experimental observations. 
Efforts in this direction are already on \cite{Conley:2010du,Conley:2011nn}.
In fact, in some of their recent studies, the ATLAS and CMS collaborations 
had also presented exclusion plots in a framework like the much relaxed 
phenomenological MSSM \cite{atlas-pub-3,atlas-pub-b1, atlas-prel-b1, 
atlas-prel-b2, cms-prel-2} and also referred to the so-called \emph{simplified} 
models \cite{atlas-pub-2,atlas-pub-5,cms-pub-2} following 
Refs.\cite{Alwall:2008ag,Alves:2011sq,Alves:2011wf}.

Under the circumstances, the SUSY partners of the third generation quarks 
(the scalar quark or squarks) have got a special standing from a few 
different but somewhat compelling considerations. For example, it is well 
known that the squarks from the third generation, \emph{viz.}, the top 
squark(s) or the `stop' ($\tilde{t}$) and the bottom squarks(s) or the 
sbottom ($\tilde{b}$), may be rather light (compared to their first two 
generation peers) due to two primary reasons. First, in a high scale scenario
where soft SUSY breaking masses are defined at a high scale, the ones
for the third generation run down to a much lower value at the weak scale
due to the large Yukawa couplings they have.
Thus, at the weak scale, the diagonal entries in the scalar mass-squared
matrices are generically smaller than those from the first two generations.
This is the so-called inverted mass hierarchy 
\cite{Baer:1994zt,Barger:1999iv,Baer:1999md,Chattopadhyay:2000qa}
where the scalar (SUSY) partners of the massive SM fermions turn out to 
have smaller soft-masses and vice versa. In addition, as electroweak symmetry
breaks, large mixing of the chiral states driven primarily 
by the mass of the partner fermions (the top and the bottom quarks) contributes
to further lowering of the mass eigenvalues for the third generation sfermions.

Adhering to a somewhat conservative interpretation of the LHC constraints, 
thus expecting, in general, super-TeV masses for the gluino and the squarks 
from the first two generations \cite{Sekmen:2011cz},
a legitimate search strategy may be to focus on the squarks from 
the third generation \cite{Desai:2009ex,Brust:2011tb}. In any case, irrespective of
what lower limits are set by an experiment on the first two generation squark 
masses, it is always reasonable to expect them to be rather heavy to satisfy
different SUSY flavour constraints while allowing for lighter third 
generation squarks with which the constraints are still much relaxed.
Also, recent dedicated studies at ATLAS and CMS on final states involving 
$b$-jets 
\cite{atlas-pub-b1,atlas-prel-b1,atlas-prel-b2,cms-pub-b1,talk-lepton-photon}
in scenarios, where only $\sbot1$ and/or 
$\sstop1$ are lighter than the gluino, tend to allow for a top or bottom 
squark much lighter than 500 GeV while for the gluino and the squarks from 
the first two generations the lower bounds are almost touching 500 GeV and 
1 TeV, respectively.

Thus, at a time when dedicated search for the squarks from the third 
generation would be on, while other strongly interacting superpartners still 
remaining unreachable, the preparedness to decipher the most from a 
positive signal would be something of genuine priority.
There have been some studies in the past \cite{Hisano:2002xq,Hisano:2003qu} 
where techniques of measuring
masses of the squarks from the third generation were discussed in scenarios
where these squarks are produced in the decays of gluino. In particular, these
studies suggested studying the edge structure of the $m_{tb}$ distribution
when gluino decays to $tb\ch1$ through a light stop or sbottom. 
Further, in recent times, phenomenology of stop as the next to lightest
supersymmetric particle (NLSP) has been considered at the LHC and/or Tevatron
in scenarios like Gauge Mediated SUSY Breaking (GMSB) \cite{Kats:2011it}
or the CMSSM \cite{Huitu:2011cp}. Interesting studies of probing the
third generation squark sector, in particular, the stop sector through
the study of the Higgs bosons were taken up in Refs.
\cite{Belanger:1999pv,Dermisek:2007fi}.

Once we know about the primary quantities like the production cross sections 
of these SUSY particles and their masses, we would turn to learn about their
couplings. These, together, would be reflective of the amount of chiral 
admixtures present in the mass eigenstates of the bottom and the top squarks. 
Dedicated study in this area is not abound, particularly in the context of 
LHC. However, a rather recent work \cite{Rolbiecki:2009hk} put the 
issue in perspective with the top squark sector in reference. Further, there 
are recent works \cite{Plehn:2010st,Plehn:2011tf} proposing new techniques 
to reconstruct the top squark by tagging the top quarks.

As we will see in the next section, the observables in the sbottom and the
stop sectors, particularly the basic ones like the masses and the mixings, 
are controlled by some parameters which are common to both. Incidentally,
some of them also control the compositions of the charginos and the 
neutralinos in an important way. Thus, it is only natural to expect some 
reasonable correlations inherent in these sectors. Such correlations might 
provide some phenomenological handle in systematic explorations of these 
sectors in tandem and this is what we like to address here.

In this work, we perform a case-study with the production of the lighter 
bottom squark at the 14 TeV LHC which eventually decays either to a top 
squark and a SM $W^\pm$ boson or to a lighter chargino and a top quark or 
to a bottom quark along with a neutralino. Note that the first of these 
decay modes is the only one that involves squarks from both sbottom and the
stop sectors and thus could reflect on the simultaneous features of these
two sectors, i.e., the so-called entanglement. Note that the decay
$\sbot1 \to \sstop1 W^+$ boson would be significantly enhanced only if 
both of the squarks have significant $SU(2)$ (left) admixtures. This decay 
mode was first discussed in Ref.\cite{Hidaka:2000cm}. In this work, we point out 
that this is where the entanglement between the sbottom and the stop 
sectors is likely to play an important role. 

For all the decay modes 
mentioned above, the resulting final states would contain $b$-jets; either 
from the direct decays of the bottom squark or from the subsequent decays of 
the top squark or the top quark. It was also noted in Ref.\cite{Hidaka:2000cm} 
that the $p_T$ spectrum of the $b$ quarks, the $W^\pm$-bosons and the missing 
$p_T$ spectra could be different for the above decay modes. Thus, the count of 
events enriched with bottom quarks, $W^\pm$ bosons and missing $p_T$ with 
characteristic kinematic properties could be indicative of the mixing-pattern 
present in the sbottom and stop sectors. We also point out that the LHC 
running at 7 TeV is not likely to offer much insight in such a study.

he paper is organised as follows. In section 2, we briefly outline the 
roles of different SUSY parameters that control the masses and the mixings 
in the sbottom and the stop sectors. We indicate how some of these 
parameters affect the compositions of the charginos and the neutralinos. 
We also touch upon the kind of correlation these common parameters
might bring into the combined system. In section 3, we discuss the signals 
and the potential backgrounds at the LHC. A brief account of the setup and 
our analysis followed by the important observations are presented in section 4.
In section 5, we summarise.
%

\section{Masses and mixings}
In this section, we outline the correlated nature of the system comprised 
of the third generation squarks and, to a varied extent, the charginos and 
the neutralinos. 

The mass-squared matrix for the squarks, more relevantly, those from the 
third generation has the following generic form in the 
($m_{\widetilde{Q}_{_{3L}}}$, $ m_{\tilde{q}_{_{3R}}}$) basis:
\begin{equation}
M_{ {\tilde{q}_{_3}}}^2 =  
\left(
\begin{array}{c c}
m_{LL}^2 + m_{q_{_3}}^2  & m_{q_{_3}} m_{LR}\\
 m_{q_{_3}} m_{LR} &  m_{RR}^2 + m_{q_{_3}}^2
\end{array}
\right )
\end{equation}
\noindent
where
\begin{eqnarray}
m_{LL}^2 &=& m^2_{\widetilde{Q}_{_{3L}}}+ (T_3-e \sin^2 \theta_W) 
\, m_Z^2 \cos 2\beta  \nonumber \\
m_{RR}^2 &=& m^2_{\tilde{q}_{_{3R}}}   + e \sin^2{\theta_W} 
\, m_Z^2 \cos 2\beta  \nonumber \\
m_{LR}^2 &=& A_{q_{_3}} - \mu R
\end{eqnarray}
\noindent
where $m_{q_{_3}}$ is the bottom or the top quark mass, 
$m_{\widetilde{Q}_{_{3L}}}$ ($ m_{\tilde{q}_{_{3R}}}$) is the 
SUSY-breaking soft mass for the left (right)-chiral bottom or the 
top squark, $T_3$ and
$e$ are the isospin quantum number and the electric charge of the squark
in context, $\theta_W$ is the Weinberg angle and 
$A_{q_{_3}}$ is the soft SUSY breaking scalar trilinear parameter that appears
in the scalar potential of the MSSM ($A_b$ or $A_t$, as the case may be).
$R$ is equal to $\tan\beta$ ($\cot\beta$) for the sbottom (stop) sector.
The second term in each of the first two expressions of equation (2) is
proportional to $m_Z^2 \cos 2\beta$ and are known as the $SU(2)_L$ 
and $U(1)_Y$ $D$-terms originating in the quartic term of the scalar potential. 
%

Further, let us write down the general formula for the mass eigenvalues 
of the physical states obtained on mixing of the chiral degrees of freedom:
\begin{equation}
m^2_{({\tilde{b},\tilde{t})}_{(1,2)}} =
{1\over 2} \left(2 m_{b,t}^2 + m_{LL}^2 + m_{RR}^2 \mp
\sqrt{(m_{LL}^2-m_{RR}^2)^2 + 4 m_{LR}^2 m_{b,t}^2} \right)
\end{equation}
where the subscript 1(2) on the left hand side corresponds to the 
negative (positive) sign in front of the term under square-root on the
right thus ensuring the mass-variable with subscript 1 (2) is the lighter 
(heavier) of the two mass eigenstates.
\noindent
The expressions involving the mixing angles are as follows:
\begin{eqnarray}
\cos\theta_{\tilde{b},\tilde{t}} &=&
{{-m_{b,t} \, m_{LR}} \over \sqrt{m_{b,t}^2 m_{LR}^2 
+ (m^2_{ {\tilde{b}_1,\tilde{t}_1} }-m_{LL}^2)^2}} \nonumber \\
\sin\theta_{\tilde{b},\tilde{t}} &=&
{{m_{LL}^2 -m^2_{ {\tilde{b}_1,\tilde{t}_1} } }  \over \sqrt{m_{b,t}^2 m_{LR}^2 
+ (m^2_{ {\tilde{b}_1,\tilde{t}_1} }-m_{LL}^2)^2}}.
\end{eqnarray}
The lighter mass eigenstates in terms of the mixing angles have the
general form:
\begin{eqnarray}
\sstop1 &=& \cos\theta_{\tilde{t}} \stopl + \sin\theta_{\tilde{t}} \stopr 
\nonumber \\
\sbot1  &=& \cos\theta_{\tilde{b}} \sbotl + \sin\theta_{\tilde{b}} \sbotr
\end{eqnarray}
while the heavier ones are orthogonal states, respectively.
\noindent
Note that $0 < \theta_{\tilde{b},\tilde{t}} < \pi$ by convention.
Also, under this convention,  a mass eigenstate is purely left-chiral
in nature for $\theta_{\tilde{b}, \tilde{t}} =0^\circ$ and entirely
right-chiral when $\theta_{\tilde{b}, \tilde{t}} =90^\circ$.
A detailed description of this sector can be found in the thesis work
of Ref.\cite{Kraml:1999qd}.

A key issue which we like to exploit here in this work is that the soft 
mass for the left-chiral degrees 
of freedom ($m_{\widetilde{Q}_{_{3L}}}$) is common to both sbottom and 
stop sectors. The sole difference between the two sectors in this part 
arises due to the so-called $D$-term contributions which are different
for the two members of the isospin-doublet ($\mstopl$ and $\msbotl$). 
On the other hand, $m_{\tilde{q}_{_{3R}}}$ for the two sectors can be 
different ($\mstopr$ and $\msbotr$, respectively) as these correspond 
to isospin singlets not related by any symmetry.

A few important points pertinent to the present study that emerge from equations 
(1), (2) and (3) are as follows. 
Clearly, through its presence in the off-diagonal terms 
of the mass-squared matrices, $\mu$ controls the mixings, \emph{i.e.}, the 
chiral contents of the mass eigenstates, provided the trilinear couplings 
are not too large. Also, note that $\mu$ appears as a product with 
$\tan\beta$ (in the sbottom sector) or $\cot\beta$ (in the stop sector). 
Hence, choice of $\tan\beta$ 
as a basic parameter could determine the role of $\mu$ by tempering the 
product-term up to an order of magnitude over the range $5 < \tan\beta < 50$.
The patterns of such mixings are effectively captured, in a somewhat global 
way, in the parametrisation of Ref.\cite{Demina:1999ty}  in terms of a ratio 
of the diagonal and the off-diagonal entries in the mass-square matrix.
It can be easily checked that for 
$A_t \, (A_b) << \mu \cot\beta \, (\mu\tan\beta)$, 
the off-diagonal terms in the two sectors are comparable when 
$\tan\beta \approx 6$ which is uniquely determined by the ratio $\mtop/\mbot$. 
Thus, in a situation where the diagonal terms of the mass matrices for these 
two sectors are comparable, the role of $\mu$ in determining the mixing would 
be similar under such circumstances, for any value of $\mu$.

Further, it is to be noted that the relative values of the soft masses 
$m_{\widetilde{Q}_{_{3L}}}$ and $m_{\tilde{q}_{_{3R}}}$ appearing in the 
diagonal terms of the mass-squared matrix influence the chiral contents 
of the sbottom and the stop mass eigenstates in a definitive way in the
presence of mixing. For example, maximal mixing in the top (bottom) squark 
sector requires $\mstopl (\sbotl) \approx \mstopr (\msbotr)$ at the weak 
scale. This is a robust, but rather intuitive, requirement. Larger 
the off-diagonal term is, larger a deviation from the equality of these
masses can be afforded for obtaining maximal mixing. Thus, assuming that 
the off-diagonal terms of the mass-squared matrices in both the sectors 
are likely to be primarily controlled by the corresponding fermion masses, 
almost near equality of the soft chiral masses in the sbottom sector is 
required to yield close to maximal mixing. By the same token, for the stop 
sector, a somewhat larger difference in the chiral soft masses can still 
result in a large mixing for given values of $A_{b,t}$, $\mu$ and 
$\tan\beta$. 

However, such near equalities are \emph{unlikely} to be achieved in 
generic high scale scenarios with universal scalar masses (including 
mSUGRA) and with a grand 
desert between the high scale and the electroweak scale. Take for example, 
an $SO(10)$ GUT inspired scenario with a universal scalar mass 
$m_0$ \cite{Kawamura:1994ys} where the $SO(10)$ breaks directly 
to the SM gauge group at the GUT scale. Both left and right top 
squarks leave in the same $5$-plet of the $SU(5)$ embedded in 
$SO(10)$. This would lead to (up to effects from physics above 
the GUT scale) $\mstopl= \mstopr$ at the GUT scale. It immediately 
follows that  $\mstopr < \mstopl$ at the weak scale due to 
renormalisation group (RG) running. In contrast, $\msbotr$ can be 
different from $\msbotl$ at the high scale since $\sbotl$ and 
$\sbotr$ reside in different multiplets of $SO(10)$. Hence, at the 
weak scale, $\msbotl$ and $\msbotr$ can have any relative hierarchy 
if nonuniversality in scalar masses at some high scale is allowed for.
Thus, in particular GUT motivated scenarios, the chiral contents of the 
sbottom mass eigenstates may vary in a more relaxed fashion unlike 
those of the stop mass eigenstates.

Thus, while maximal mixing in the sbottom sector can somewhat more 
naturally be achieved by fulfilling the requirement $\msbotl \approx \msbotr$ , 
the same in the top squark sector can hardly be realized in popular GUT 
motivated scenarios that assume universal scalar masses at the high
(unification) scale and the presence of a `grand desert' between the 
latter and the weak scale. In other words, observation of 
maximal mixing in the stop sector could very well indicate a departure 
from this popular paradigm. In this work, we try to exploit this issue. 
Thus, we adopt a purely phenomenological approach. We vary the chiral 
soft masses in the stop sector freely to explore the mixing in this 
sector and its implications. 

On the other side of the story, the values of the soft masses of the
$U(1)$ and $SU(2)$ gauginos ($M_1$ and $M_2$), with respect to $\mu$, 
are crucial for the masses and the compositions of the charginos and 
the neutralinos (their gaugino and higgsino contents) which, in turn, 
broadly dictate the phenomenology involving them. Note that the
charginos and the neutralinos interact with sbottom and stop eigenstates
with both gaugino and higgsino components they have. This immediately 
hints towards a possible bridge that $\mu$ could provide between the two sectors. 
In addition, the role of $\mu$ can be seen in conjunction with that of 
$\tan\beta$ in the stop and sbottom sector (as explained above) while 
the role of the latter in the chargino and the neutralino sectors could 
assume importance under specific situations. An illustrative study to this 
end encompassing both the sectors is beyond the scope of the present work 
and will be taken up in a subsequent study \cite{future-work}.

In this work we restrict ourselves to the study of possible correlations 
that might be present in the sbottom and the stop sectors. 
Such correlations are likely to be best manifested in studies where both 
sectors have explicit involvements. The most significant of such situations 
could be realised in the productions of the lighter sbottom (lighter stop) 
at the colliders followed by its decays to the stop (sbottom) where 
$W^\pm$-bosons appear. We thus focus on a particular case of production of 
bottom squarks at the LHC followed by their subsequent decays with an aim 
to study the imprints of mixings in the involved sectors.
%

\section{The Signal and the Background}
In this section we present some aspects of phenomenology of the lighter 
of the bottom squarks at the LHC. Recently, in the context of a SUSY/MSSM
`golden region' \cite{Perelstein:2007nx,Perelstein:2007st}, such a study 
was taken up in Ref.\cite{Li:2010zv}. This was necessarily limited, though 
in a rather motivated way, to a low $\mu$ regime with all the squarks from 
the first two generations taken to be rather heavy. The study stuck to an
appropriate final state. In the present study, for our purpose, we adopt a 
rather open approach and analyse other possible final states originating 
from sbottom production at the LHC. This, as we will see, would inevitably 
involve the top squarks and thus may potentially shed light into the chiral 
compositions of the bottom and the top squarks in a correlated way. As 
discussed in the Introduction, this might even have reference to the 
texture of the gaugino sector in the same go.

Unlike in Ref.\cite{Li:2010zv} where only pair-production of bottom 
squarks was considered, we keep the options open for other production
processes which may eventually lead to a pair of bottom 
squarks at the LHC. 
These include production of a lighter bottom squark in association with 
a gluino and pair-production of gluinos where a gluino may decay 
subsequently to a bottom quark and an sbottom. These three modes of
lighter sbottom production are shown schematically below:
\begin{center}
\begin{itemize}
\item $p p \to \sbot1 \sbotc1 \, , \, \sbot1 \sbot1 \, , \, \sbotc1 \sbotc1$ 
\item $p p \to \gluino \sbot1 \, , \, \gluino \sbotc1 \; : \; 
       \gluino \to b \sbotc1 \, , \, \bbar \sbot1$ 
\item $p p \to \gluino \gluino \; : \; 
       \gluino \to b \sbotc1 \, , \, \bbar \sbot1$ 
\end{itemize}
\end{center}
Now, an sbottom would always lead to a bottom quark in its decay, be it 
from a direct decay or through top quark production under a cascade.
These are illustrated below.
\begin{center}
$\bullet$  $\sbot1 \to b \chi^0_{i=1-4}$ 
\hskip 20pt $\bullet$ $\sbot1 \to t \chi^-_{j=1,2}$
\hskip 20pt $\bullet$ $\sbot1 \to \sstop1 W^-$ 
: $\sstop1 \to b \chjp \, , \, t \ntrli$
\end{center}
Note that any of these modes could lead to at least a pair of lighter bottom 
squarks and up to four bottom quarks (in the case of gluino pair-production) 
in the final state. 

As indicated above, the bottom squark may have different possible decay modes 
out of which the ones to $\sstop1 W^\pm$, $t\chpm1$, $b\ntrl1$ and $b \ntrl2$
would be of importance. As for $\sstop1$, its decays to $t\ntrl1$, $t\ntrl2$, 
$b\chpm1$ would be in context. Further, the gauge bosons (like $W^\pm$), the 
charginos and the neutralinos all would contribute to both leptonic and 
hadronic final states. In fact, as pointed out in the Introduction, the decay 
channels $\sbot1 \to \sstop1 W^\pm$ and $\sbot1 \to t\chpm1$ could lead to 
identical final states. Hence, if the branchings in these two channels are 
complementary in nature, 
rates for the final state events may not be much sensitive to the variation 
of the same. Thus, they may not shed much light on the couplings (and 
hence, on the mixings angles) involved. We will get back to this issue in 
the next section where we study situations under which clearer imprints of 
branchings and hence mixing angles are left in 
the events rates of the different final states. We would also contrast those 
to a generic scenario where all decay modes of $\sbot1$ are open and some of 
them are significant.

It is clear from the above discussion that the multiplicities of bottom 
quarks in the final state, at the parton-level, may vary between 2 and 4. 
While an experimental study of final states even with a moderate bottom 
quark multiplicity is a challenging proposition, their presence, nonetheless,
would boost the counts for low-multiplicity (up to 2) final states through 
enhanced combinatoric factors.
With these general possibilities in mind, we pick a benchmark (reference) 
MSSM spectrum in the next section for our subsequent analysis.

The final states we consider are the following:
\begin{enumerate}
\item 2 $b$-\emph{jets} + $\ge 4$ \emph{jets} + $\ge$ 1-\emph{lepton} + $\etmiss$
\item  2 $b$-\emph{jets} + $\ge 4$ \emph{jets} + 
\emph{same-sign dilepton (SSDL) pair} + $\etmiss$
\item  2 $b$-\emph{jets} + $\ge 4$ \emph{jets} + 
\emph{opposite-sign dilepton (OSDL) pair} + $\etmiss$
\end{enumerate}  
where by leptons only the electrons and the muons are meant. The first of 
these final state has been studied in Ref.\cite{Li:2010zv} 
as mentioned in the beginning of this section. In our 
case, this remains to be an important channel for two reasons: first, 
because the presence of a lepton in the final state helps negotiate the 
otherwise large pure QCD background and second, because it is less 
suppressed compared to dilepton final states indicated above. 
However, as expected, for the dilepton 
final states the backgrounds are further suppressed thus making them worthy 
of a closer study. In addition, there is always the well-known advantage of 
studying signals in multiple channels which, when studied simultaneously, 
may potentially shed light on intricate issues pertaining to the 
mass-spectrum and the involved couplings in a more efficient and definitive 
way.

\vskip 10pt
As for the backgrounds for the final states indicated above, we considered
several SM processes that could be potentially strong. These are
$Zt\tbar+0,1,2,3 \;jets$, $Wt\tbar+0,1,2,3 \; jets$, $t\tbar+3,4 \; jets$,
$tbW+0,1 \; jets$. Note that $t\tbar$-pair production and the same with 
extra hard jets in low multiplicity (up to 2 extra jets) are unlikely to yield
serious background because of the minimum lepton and jet multiplicities we 
required for our signal. Fakes from charm and light quark jets are not taken 
into account in this study.
\section{The Setup and the Analysis}
The setup for our analysis is based on two `benchmark' scenarios about 
which variations are studied. These consist of two sets of weak-scale 
MSSM input parameters arranged in a minimalistic way for the purpose. 
These would be sufficient to demonstrate the goals of the present study. 
The first one is shown in Table \ref{table:high-mass}. 
This explores the prospect of a 
heavier spectrum where the masses of the gluino and that of the squarks 
from the first two generations are somewhere near or just exceeding their 
present bounds in conformity with recent studies at the LHC and their 
interpretations within the much relaxed 19-dimensional pMSSM framework 
\cite{Sekmen:2011cz}. 
The second benchmark scenario is elaborated in Table \ref{table:low-mass}. This, in turn, 
represents a lighter spectrum with somewhat lighter squarks from the 
third generation along with lighter gluino, charginos and neutralinos 
that are still very much allowed by the LHC data, as explained later in 
this section. 

In addition, for these two spectra, we ensure compatibility with other
experimental constraints like the lower bound on the mass of the lighter 
chargino obtained from the LEP experiments ($\approx 105$ GeV) and those 
pertaining to the anomalous muon magnetic moment 
($a_\mu=\frac{g_\mu-2}{2}$) \cite{Martin:2001st,Djouadi:2006be} 
and the rare Flavour Changing Neutral Current (FCNC) process like 
$b\to s\gamma$ \cite{Mahmoudi:2007vz,Arbey:2011un} 
while choosing these benchmark points. In fact, the observations in the 
latter two experiments, when considered in conjunction, favour $\mu>0$ 
\cite{Feng:2001tr}. This is why we choose to work with positive values 
of $\mu$ in the present study. As for the SM-like lightest SUSY Higgs 
boson, we require a somewhat relaxed lower bound of 111 GeV as against 
the actual LEP constraint of 114.4 GeV \cite{Barate:2003sz,Schael:2006cr}. 
This relaxed bound is consistent with the more precise estimation 
\cite{Degrassi:2002fi,Heinemeyer:2004ms} of theoretical uncertainty 
involved in predicting the mass of the Higgs boson\footnote{Strict 
compliance to the LEP Higgs bound and other precision observables like 
$a_\mu$ and BR[$b \to s \gamma$] are enforced only to these reference 
points. Everywhere else, where the purpose is to explore the sensitivity 
of the event rates to the mixing angles, we just generated the combinations 
of these angles by varying the relevant MSSM inputs without worrying about 
their compatibility with these experimental observations.}.

As indicated in the Introduction, our goal is to work with a somewhat 
light bottom squark which is expected to be 
within the reach of LHC running at its design centre of mass energy, 
i.e., 14 TeV. 
Moreover, for our purposes, such a sbottom should have enough phase space
to decay into a top squark along with a 
$W$ boson. The probability of such a decay would then be maximised when 
both $\sbot1$ and $\sstop1$ have substantial \emph{left} chiral ($SU(2)$) 
admixture. However, by requiring such an admixture 
in them, it is impossible to achieve a mass-splitting of
$\Delta m_{\tilde{b}_1 \tilde{t}_1}\sim 
\Delta m_{\tilde{b}_L \tilde{t}_L}\geq m_W$ 
between these two states. This is because, in such a limit, 
their masses at the weak scale 
are related, and at the lowest order, differ only  by an $SU(2)$ $D$-term 
which is anyway not large (see equation (2), first expression). 
Hence, a compromise is needed and chiral mixings 
are to be allowed for. This would then definitely suppress the coupling 
$\tilde{b}_1 \tilde{t}_1 W^\pm$ but at the same time open up the 
required phase space for the decay 
$\tilde{b}_1 \to \tilde{t}_1 W^\pm$ to take place.
%
\vskip 10pt
\begin{table}[htb]
\begin{center}
\begin{tabular}{|c|l|l|}
\hline 
Parameters & \hskip 50pt Input values & \hskip 32pt Output spectrum \\
\hline
Gaugino masses & $M_1=200$ & $m_{{\chi}_{_1}}^0 \approx 198$ \\
(in GeV) &  \hskip 50pt $M_2=400$  & 
\hskip 10pt $m_{\chi_{_2}^0} \approx 403$  \hskip 10pt $m_{\chi_{_1}^\pm} \approx 403$  \\
 & \hskip 100pt $M_3 = 1200$ & \hskip 90pt $m_{\tilde{g}} \approx 1193$  \\
\hline 
Chiral Squark masses &  & \\
(1st and 2nd generations) & \hskip 50pt $m_{\tilde{q}^{1,2}_{_{L,R}}}=1000$ & 
 \hskip 20pt $m_{\tilde{q}^{1,2}_{_{L,R}}} \approx 1005-1010$ \\
(in GeV) & & \\
\hline
Third generation & $m_{\tilde{Q_3}_L}=700$ & 
\hskip 0pt $m_{\tilde{b}_1} \approx 709$ 
\hskip 20pt $m_{\tilde{b}_2} \approx 1014$  \\
chiral squark masses & \hskip  50pt $m_{\tilde{b}_R}=1000$ & 
$\qquad \qquad \;\; \thetasbot =1.9^\circ$ \\
(in GeV)  & \hskip 100 pt $m_{\tilde{t}_R}=500$ &
\hskip 0pt$m_{\tilde{t}_1} \approx 455$ 
\hskip 20pt $m_{\tilde{t}_2} \approx 757$  \\
and Mixing Angles & &  $\qquad \qquad \;\;\thetastop=67^\circ$ \\
\hline
Slepton Masses  & \hskip 50pt $m_{\tilde{\ell}_{L,R}}=600$ & \\
(in GeV) &  & \hskip 50pt $m_{\tilde{\ell}} \approx 600$ \\
\hline
$A$-parameters & \hskip 25pt $A_b = 0$ &  \hskip 60pt ----- \\
(in GeV) & \hskip 75pt $A_t= -800$ & \hskip 60pt -----\\
\hline
$m_A$ (in GeV) & \hskip 70pt 500 & 
 \hskip 7pt $m_{\chi_{_3}^0}\approx 699$  \\
$\mu$ (in GeV) & \hskip 70pt 700 & \hskip 43pt $m_{\chi_{_4}^0} \approx 711$ \\
$\tan\beta$ & \hskip 75pt 10 & \hskip 80pt $m_{\chi_{_2}^\pm} \approx 711$ \\
 & & \\
\hline
\end{tabular}
\end{center}
\caption{
A somewhat \emph{heavy} benchmark SUSY spectrum and the weak-scale values of the
MSSM input parameters used to obtain the same. The input soft mass of the
CP-odd Higgs is taken to be 500 GeV. The SUSY spectrum generator used for
the purpose is Suspect v2.31. Throughout the analysis $m_{top}=172.5$ GeV 
is used. Note that $m_{\squark} < m_{\gluino}$.}
\label{table:high-mass}
\end{table}

In Table \ref{table:high-mass} we collect the set of relevant MSSM input 
parameters for the high-mass
reference point and the resulting output spectrum that conform with the 
setup described above. We call this the high-mass benchmark point/spectrum.
Note that the mixing angle in the sbottom sector, 
$\thetasbot$ is rather small thus making the lighter sbottom an almost 
pure left chiral state. However, $\thetastop$ is around 67$^\circ$ and 
hence it has some (15\%) left chiral admixture. By varying the 
relevant MSSM inputs about their benchmark values, the mixing in the top 
squark sector can be significantly altered. This would affect the decay 
branching fractions of the bottom squark. As we would see later in this 
section, this can have some impact on the signatures at the LHC.
%
\vskip 10pt
\begin{table}[htb]
\begin{center}
\begin{tabular}{|c|l|l|}
\hline 
Parameters & \hskip 50pt Input values & \hskip 25pt Output spectrum \\
\hline
Gaugino masses & $M_1=100$ & $m_{{\chi}_{_1}}^0 \approx 100$ \\
(in GeV) &  \hskip 50pt $M_2=200$  & 
\hskip 10pt $m_{\chi_{_2}^0} \approx 207$  
\hskip 10pt $m_{\chi_{_1}^\pm} \approx 207$  \\
 & \hskip 100pt $M_3 = 600$ & \hskip 90pt $m_{\tilde{g}} \approx 655$  \\
\hline 
Chiral Squark masses &  & \\
(1st and 2nd generations) & \hskip 50pt $m_{\tilde{q}^{1,2}_{_{L,R}}} = 1000$ & 
 \hskip 20pt $m_{\tilde{q}^{1,2}_{_{L,R}}} \approx 1005-1010$ \\
(in GeV) & & \\
\hline
Third generation & $m_{\tilde{Q_3}_L}=480$ & 
\hskip 0pt $m_{\tilde{b}_1} \approx 500$ 
\hskip 20pt $m_{\tilde{b}_2} \approx 715$  \\
chiral squark masses & \hskip  50pt $m_{\tilde{b}_R}=700$ & 
$\qquad \qquad \;\; \thetasbot =4.3^\circ$ \\
(in GeV)  & \hskip 100 pt $m_{\tilde{t}_R}=390$ &
\hskip 0pt$m_{\tilde{t}_1} \approx 349$ 
\hskip 20pt $m_{\tilde{t}_2} \approx 571$  \\
and Mixing Angles & &  $\qquad \qquad \;\;\thetastop = 57.3^\circ$ \\
\hline
Slepton Masses  & \hskip 50pt $m_{\tilde{\ell}_{L,R}} = 600$ & \\
(in GeV) &  & \hskip 50pt $m_{\tilde{\ell}} \approx 600$ \\
\hline
$A$-parameters & \hskip 25pt $A_b = 0$ &  \hskip 60pt ----- \\
(in GeV) & \hskip 75pt $A_t= -500$ & \hskip 60pt -----\\
\hline
$m_A$ (in GeV) & \hskip 70pt 500 & 
 \hskip 7pt $m_{\chi_{_3}^0} \approx 845$  \\
$\mu$ (in GeV) & \hskip 70pt 850 & \hskip 43pt $m_{\chi_{_4}^0} \approx 849$ \\
 $\tan\beta$ & \hskip 75pt 10 & \hskip 80pt $m_{\chi_{_2}^\pm} \approx 850$ \\
 & & \\
\hline
\end{tabular}
\end{center}
\caption{
Same as in Table \ref{table:high-mass} except for a somewhat \emph{smaller} 
masses for the lighter charginos and neutralinos, the gluino and the third 
generation squarks. Note that $m_{\squark} > m_{\gluino}$.
}
\label{table:low-mass}
\end{table}

To decide on the benchmark spectrum with lighter masses 
(Table \ref{table:low-mass}), 
a closer look at recent LHC studies \cite{atlas-pub-b1,atlas-prel-b1,
atlas-prel-b2, talk-lepton-photon} is warranted. These analyses discuss 
signals with heavy flavour jets ($b$-jets) and large missing transverse 
energy without \cite{atlas-prel-b1} and with \cite{atlas-prel-b2} an 
isolated lepton ($e$ and/or $\mu$). There, either $\sbot1$ 
\cite{atlas-prel-b1} or $\sstop1$ \cite{atlas-prel-b2} are assumed 
to be the lightest squark and the branching fractions of gluino decaying 
into them (in the respective cases) is 100\%. Thus, in these studies, 
$\sbot1$-s or $\sstop1$-s are produced either directly in pairs or via 
production and subsequent decays of gluino. 
Further, it is assumed in these experimental studies that $\sbot1$ and 
$\sstop1$ always decay in specific channels, i.e., BR[$\sbot1 \to b \ntrl1$] 
\cite{atlas-prel-b1} and BR[$\sstop1 \to b \chp1$] \cite{atlas-prel-b2} 
are 100\% in the respective cases. Ref.\cite{atlas-prel-b1} rules out
$\mgluino < 720$ GeV for $\msbot1 < 600$ GeV. Ref.\cite{atlas-prel-b2} 
excludes $\mgluino < 500-520$ GeV for $\mstop1$ between 125 GeV and 300 GeV. 
In Ref.\cite{atlas-prel-b2}, exclusion limit is also placed on the
$\mgluino-\mntrl1$ plane when $\sstop1$ is still the lightest of the squarks 
but is heavy enough ($\mgluino < \mstop1 + m_t$) such that $\gluino$ cannot 
decay into an on-shell $\sstop1$ and it decays via off-shell 
$\sstop1$ to $t\tbar \ntrl1$ final state with a decay branching fraction of 
100\%. The analysis excluded $\mntrl1 < 40(80)$ GeV for 
$\mgluino < 570 (540)$ GeV. All these reported exclusions are at 95\% 
confidence level.

However, the above limits may be considered conservative
because they assume the Br[$\gluino \to \sbot1 \bbar \, + \, h.c.$] or 
Br[$\gluino \to \sstop1 \tbar \, + \, h.c.$]
to be 100\%. In a situation where there is more than one third-generation 
squark lighter than the gluino, this assumption would not hold and may result
in relaxed bounds.
In addition, note that events from direct stop pair production are reported 
\cite{atlas-prel-b2} to have a very low acceptance; presumably due to hard 
kinematic cuts used in the analysis. These issues may very well
dilute the sensitivities of these experimental analyses (to a given data-set) 
which would eventually lower the exclusion limits. In Table
\ref{table:low-mass}, we tried to exploit this caveat to our advantage
and settled on a somewhat lower mass for the gluino ($\approx 655$ GeV) 
in relation to $\msbot1$ ($\approx 500$ GeV) considered.

For the two benchmark scenarios, we work in an otherwise unconstrained SUSY 
scenario except retaining an imprint of unification of gaugino masses (at a 
high scale) in the choice of their weak scale values. Note, however, that such 
a choice is only to keep the analysis tractable and does not feature an 
essential part of our study. As is well-known, departure from such an 
assumption, can easily have nontrivial phenomenological implications. Two such 
examples are discussed in the context of the benchmark scenarios, the emphasis 
being on the subtle handles these may provide in the analysis and interpretation
of the collider signals.  A systematic study of the implications of such 
departures is beyond the scope of the present study though and would be taken 
up elsewhere \cite{future-work}.

In Table \ref{table:xsec} we collect the lowest order cross sections for the 
strong-production processes that lead to a pair of $\sbot1$, \emph{viz.},
$pp \longrightarrow \sbot1 \sbotc1, \sbot1 \sbot1 + h.c., \; 
\sbot1 \gluino + h.c., \; \gluino\gluino$ for the 14 TeV run of the LHC. 
The cross sections are calculated by the event generator Pythia v6.420
\cite{Sjostrand:2006za} and cross-checked with CalcHEP v2.5.4 
\cite{Pukhov:2004ca}. Note that this is a conservative estimate since the 
next to leading order (NLO) contribution from QCD to squark 
(including stop and sbottom) and gluino productions 
\cite{Beenakker:1996ch,Beenakker:1996ed,Beenakker:1997ut} and
the same combined with next to leading log (NLL) resummed soft-gluon contribution 
\cite{Beenakker:2009ha,Beenakker:2010nq} may increase the cross sections by$\sim 30-40\%$ 
on an average, for our benchmark scenarios, the NLL contributions being far more
important for the case of gluino and squarks from the first two generations.
Two sets of cross sections are presented for the two benchmark spectra of Table 
\ref{table:high-mass} and Table \ref{table:low-mass}. As expected, the
respective cross sections are much larger for the lighter spectrum of
Table \ref{table:low-mass}. Also, in both cases the relative magnitudes of 
$\gluino$-pair cross section are significant. So, if there is a reasonable 
branching fraction for the decay $\gluino \to \sbot1 \bbar+h.c.$, pair-production 
of gluino could become a useful source of $\sbot1$-pair 
\cite{Hisano:2002xq,Hisano:2003qu}.


\begin{table}[t]
\begin{center}
\begin{tabular}{|c|c|c|c|c|c|c|}
\hline
Spectrum & $\sigma_{\sbot1\sbot1}$ & $\sigma_{\sbot1\gluino}$ & 
$\sigma_{\gluino \gluino}$ \\
 & (pb) & (pb) & (pb) \\
\hline
 & & & \\
Table \ref{table:high-mass} & 0.030 & 0.003 & 0.022 \\
(${\msquark}_{_{1,2}} < \mgluino$) & & & \\
 & & & \\
\hline
 & & & \\
Table \ref{table:low-mass}  & 0.265 & 0.110 & 1.764  \\
(${\msquark}_{_{1,2}} > \mgluino$) & & & \\
 & & & \\
\hline
\end{tabular}
\end{center}
\caption{Lowest order production cross sections at the LHC ($\sqrt{s}=14$ TeV) for 
different strong production processes that lead to pair of lighter bottom 
squarks using the mass-spectra of Tables 1 and 2. CTEQ6L parton 
distribution function is used with renormalisation/factorisation scale 
set at $\sqrt{\hat{s}}$.} 

\label{table:xsec}
\end{table} 

\begin{table}[htb]
\begin{center}
\begin{tabular}{|c|r|r|r|}
\hline
 & Branching fractions & Branching fractions  & Branching fractions \\
 Spectrum & $\squark$ and/or $\gluino$-decay & $\sbot1$-decay & $\sstop1$-decay \\
  & (in \%) & (in \%) & (in \%) \\
\hline
 & & & \\
 &   $\gluino \to b \sbotc1+\bbar \sbot1$ : 9.8   
 & $\sbot1 \to W^- \sstop1$ : 64.1 
 &  \\
Table \ref{table:high-mass}  & $\gluino \to t \sstopc1+\tbar \sstop1$ : 13.1  
 & $\sbot1 \to t \chm1$ : 19.6 
 & $\sstop1 \to t \ntrl1$ : 87.0 \\
(heavier masses) &   $\gluino \to q \squarkc+\qbar \squark$ : 77.1
 & $\sbot1 \to b \ntrl2$ : 15.1  
 & $\sstop1 \to b \chp1$ : 13.0 \\
$m_{\gluino}> m_{{\squark}_{_{1,2}}} > \msbot1 > \mstop1$ &  
 &   $\sbot1 \to b \ntrl1$ : 1.2  
 & \\
 & & & \\
\hline
 & & & \\
 &  $\squark_R \to q \gluino$ : 95.0  & $\sbot1 \to W^- \sstop1$ : 34.5 
 &   \\
 Table \ref{table:low-mass} &  $\squark_L \to q \gluino$ : 70.0  & $\sbot1 \to t \chm1$ : 36.2
 &   $\sstop1 \to t \ntrl1$ : 39.2 \\
 (lighter masses) &  $\gluino \to b \sbotc1+\bbar \sbot1$ : 48.5   & $\sbot1 \to b \ntrl2$ : 
27.7
 & $\sstop1 \to b \chp1$ : 60.8 \\
$m_{{\squark}_{_{1,2}}}>m_{\gluino} > \msbot1>\mstop1$ & $\gluino \to t \sstopc1+\tbar \sstop1$ : 51.5 &  $\sbot1 \to b \ntrl1$ : 1.6 & \\
 & & & \\
\hline
\end{tabular}
\end{center}
\caption{
Branching fractions of $\gluino$, $\sbot1$
and $\sstop1$ for the mass-spectra given in Tables 1 and 2. 
}
\label{table:branchings}
\end{table}

%

%
The high-mass scenario (of Table \ref{table:high-mass}) with 
$m_{{\tilde{q}}_{_{_{1,2}}}} < \mgluino$ turns out to be a rather 
conservative one. First and foremost, the individual cross sections for 
all the contributing processes are on the smaller side because the 
sparticles are heavier. Second, the branching ratio for 
$\gluino \to \sbot1 b$ is affected since 
$\gluino \to \tilde{q}_{_{_{1,2}}} q \, (+ \, h.c.)$ is kinematically 
accessible and when summed over 4 flavours, would become dominant. 
Third, we miss out on potential contributions from the production of 
$\tilde{q}_{_{_{1,2}}}$, in pair or in association with a gluino, which 
could have led up to 4 bottom quarks had the hierarchy been 
$\mstop1 < \msbot1 <\mgluino < m_{\tilde{q}_{_{_{1,2}}}}\sim \msbot2 
\sim \mstop2$.


The suppression of Br[$\gluino \to \sbot1 b$ + h.c.] is demonstrated in the 
upper part of Table \ref{table:branchings} for the heavy spectrum of Table 
\ref{table:high-mass}. It is, however, a straight-forward exercise to find 
out to what extent Br[$\gluino \to \sbot1 b$ + h.c.] may get enhanced if the 
mass-hierarchy of the gluino and the squarks from the first two generations 
is flipped, other parameter remaining the same. Note that the ratio of the 
branching fractions of the gluino to $\sbot1$ and $\sstop1$ 
(i.e., $9.8\,:\,13.1$) reflects the the ratio of the decay widths of the 
gluino in these two modes. Thus, when only these two modes add up to 100\% 
of the branching fraction, Br[$\gluino \to \sbot1 \bbar + h.c.$] would be 
around 43.5\%. More importantly, such tweaking of the squark-masses would 
immediately bring in further contributions from heavier squarks when they 
decay into gluino followed by the latter decaying into $\sbot1$. 
%
In contrast, our choice of low-mass spectrum of Table \ref{table:low-mass}
represents a favourable scenario on all respective counts laid down above.
It is evident from the lower parts of Tables \ref{table:xsec} and 
\ref{table:branchings} that the low-mass benchmark spectrum boosts both cross 
sections and its branching fractions relevant for our purpose.
Thus, the two benchmark scenarios are arranged to demonstrate the extremal 
situations.

Table \ref{table:branchings} also collects the branching fractions of $\sbot1$
and $\sstop1$ to different modes that are instrumental for the study we
undertake. For the high-mass spectrum, we see that the crucial mode
$\sbot1 \to W^- \sstop1$ has a reasonable branching fraction of about 64\%.
For the low-mass spectrum, this branching fraction drops to around 34\%,
presumably due to a combined effect of diminished splitting between 
$\sbot1$ and $\sstop1$ and possible enhancements of the competing modes.
This is a plausible explanation since, as can be seen from Tables
\ref{table:high-mass} and \ref{table:low-mass}, the left ($SU(2)$) admixtures 
in $\sbot1$ and $\sstop1$ are not different in these two cases.

One of the goals of this work is to understand how 
sensitive the event rates in different leptonic final states 
are to the mixing angles in the third generation squark sector.
As pointed 
out earlier, since all the relevant leptonic final states get 
contributions from almost all possible branching modes in this sector, 
our ability to filter out the individual contributions to the
extent possible would help extract information on 
the mixing angles involved. 

Towards this end one needs to know how sensitive the individual branching
fractions of the bottom and the top squarks are to the variations of the
mixing angles $\thetasbot$ and $\thetastop$. We choose to study the effect 
of variation of $\thetastop$ only. The reason behind this is two-fold.
First, $\thetastop$ is more sensitive to small variations in the SUSY 
inputs than $\thetasbot$ thanks to the amplifying SM factor $\mtop$ sitting
in the off-diagonal term of the corresponding mass-squared matrix. Second,
$\thetastop$ variation directly affects the branching fraction of $\sstop1$. 
Nonetheless, an analogous study with respect to $\thetasbot$
would have its own characteristics though effecting a comparable variation 
on it would require major tweaking of the soft parameters of the bottom
squark sector.

At this stage, for such a study to be meaningful, it is to be 
necessarily assumed that the masses of the lighter sbottom and the lighter 
stop are already known from the experiments. This ensures that the concerned 
kinematics remain mostly unaltered for the relevant production and decay 
processes. 
Thus, we vary $\mstopr$ and $A_t$ to achieve the variation in $\thetastop$ 
while allowing $\mstop1$ to vary only within a certain range (($\pm 10$ GeV)
about the reference value of $\mstop1=455$ GeV 
(see Table \ref{table:high-mass}). 
Clearly, such variations do not at all alter either the mass of the 
lighter bottom squark or the mixing angle in that sector.
We also keep $\mu$, $\tan\beta$, $M_1$ and $M_2$ fixed during this variation 
so that the masses and the mixings in the chargino and the neutralino sectors 
remain almost fixed in the process. Altogether, this ensures
that the variations we see in the branching fractions of the bottom and the top
squarks are almost entirely due to the variation in $\thetastop$. This would 
definitely help obtain a clearer picture of the role of $\thetastop$ in 
shaping the pattern of cascade decays of the lighter sbottom. 

In Figure \ref{figure:theta-br}, we present the variation of branching 
fractions of both $\sbot1$
and $\sstop1$ as functions of $\thetastop$. Note that the products
of different branching fractions of these excitations determine the 
predominance of certain cascade-patterns when they decay. In the
present case, the possible decay mode of lighter sbottom to lighter stop 
and charged Higgs boson is closed, the latter being rather heavy 
(resulting from our choice of an input $m_A$ of 500 GeV).
\begin{figure}[t]
\centering
\vspace*{-2.2in}
{\label{fig:theta-br}
\hspace*{-1.5cm} \includegraphics[height=30cm,width=18cm,angle=0]
{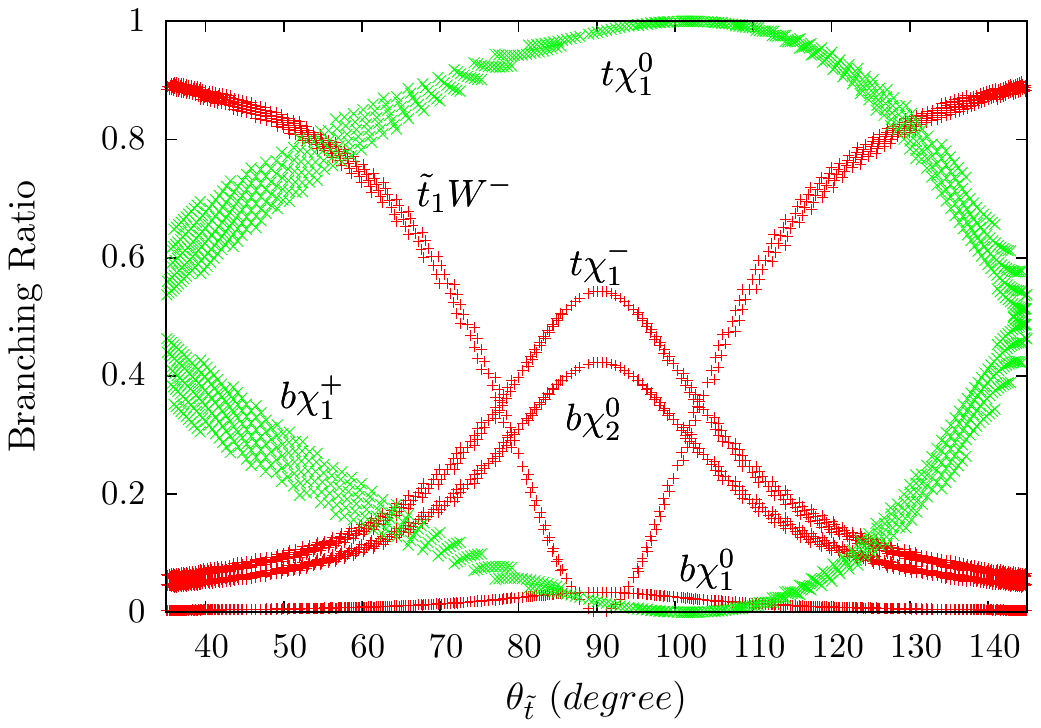}}
\vspace*{-7.0in}
\caption{\small
(a) Variations of branching fractions of the lighter bottom 
squark (red/dark grey band) and lighter top squark into 
different major decay modes as a function of $\theta_{\tilde{t}}$ while 
keeping $m_{\tilde{t}_1}$ within the range 445-465 GeV. The variation of
$\theta_{\tilde{t}}$ is achieved (see Figure \ref{figure:mtr-At-theta})
by varying the soft parameters
$m_{\tilde{t}_R}$ and $A_t$ while keeping all other parameters relevant
to the third generation squarks fixed at the benchmark values.
}
\label{figure:theta-br}
\end{figure}

Out of the decay widths that enter the calculation of the branching 
fractions of the lighter sbottom, only $\Gamma (\sbot1 \to \sstop1 W^-)$ 
depends upon $\thetastop$ and is actually proportional to 
$\cos^2\thetasbot \cos^2\thetastop$ \cite{Hidaka:2000cm}.
Thus, as a function of $\thetastop$, the variations of different 
sbottom branching fractions are solely determined by the branching 
profile of $\sbot1 \to \sstop1 W^-$ which goes as $\cos^2 \thetastop$
(and hence the symmetry about $\thetastop=90^\circ$), $\thetasbot$ being 
held fixed, as is the case here\footnote{In fact, while the
left branch with $\thetastop < 90^\circ$ arises for $A_t<0$, the right one 
with $\thetastop > 90^\circ$ results from $A_t>0$. Also, moving away on 
both sides from $\thetastop=90^\circ$, i.e., increasing amount of mixing, 
corresponds to increasing $|A_t|$. Since, with increasing $|A_t|$, mixing
is dominantly determined by $A_t$, it is expected that
further one is from $\thetastop=90^\circ$ on either side of the curve,
it is more likely that similar values of $|A_t|$ result in similar
values of branching fractions.}. 
Note that such a symmetry with respect 
to $\thetastop$ is not there in the variation of branching fractions 
of $\sstop1$. This is because the they have a somewhat complicated 
dependence on $\thetastop$. Clearly, this variation neither alters 
$\msbot1$ nor $\thetasbot$. Thus, $\sbot1$ remains to be almost purely
left chiral.

It is to be noted that the variations in Figure \ref{figure:theta-br} are 
in the form of bands. The reason behind this is that the variation of 
$\thetastop$ is achieved by simultaneous variations of  $\mstopr$ and $A_t$
over the ranges 
$300 \; {\mathrm {GeV}} \leq \mstopr \leq 1500 \; {\mathrm {GeV}}$ and 
$-3 \; {\mathrm {TeV}} \leq A_t \leq +3 \; {\mathrm {TeV}}$, respectively 
while ensuring $\mstop1$ to be roughly in the range 
$445 \leq \mstop1 \leq 465 $ GeV. 
%
\begin{figure}[htb]
\centering
\vspace*{-2.4in}
{\label{fig:theta-br}
\hspace*{-1.5cm} \includegraphics[height=30cm,width=20cm,angle=0]
{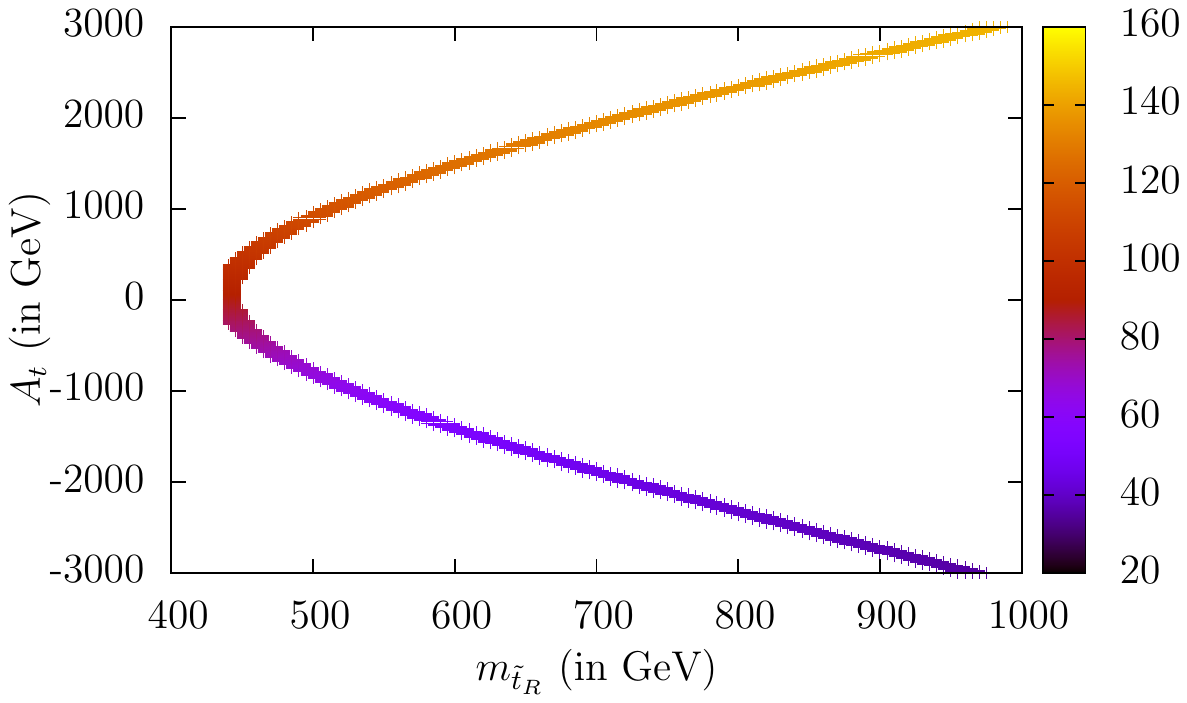}}
\vspace*{-7.0in}
\caption{\small
Ranges of $\thetastop$ in degrees presented in colour-code 
(for the setup in Figure \ref{figure:theta-br})
as $\mstopr$ and $A_t$ vary keeping 445 GeV $< \mstop1 <$ 465 GeV.
$\msq3l$ and $\msbotr$ are kept fixed at 700 GeV and 1000 GeV, respectively.}
\label{figure:mtr-At-theta}
\end{figure}

Figure \ref{figure:mtr-At-theta} reflects some important features
relevant for the phenomenology under the given setup. These are:
\begin{itemize}
\item The graph is almost symmetric about $A_t=0$. This is expected
when $A_t$ dominates in the off-diagonal term of the stop mass-squared
matrix. Given that $\tan\beta=10$ and $\mu=700$ GeV for our case and 
the off-diagonal term dominates over most of the $A_t$-range studied above.
\item The above issue is equivalent to having complementary angles and 
this is clearly seen in the figure.
\item The resulting band is rather narrow. This is because we require
$\mstop1$ be within the range 445 GeV $< \mstop1 <$ 465 GeV. 
\item For values of $\mstopr$  much larger than $\msq3l$ one requires
large values of $|A_t|$ for bringing down $\mstop1$ to the mentioned
range. The resulting mixing can be close to maximal or lower but never
reaches very small values (corresponding to dominance of left-chiral
admixture). 
\item Maximal mixings (around 45$^\circ$ and 135$^\circ$) are expected
for $\mstopr$ in the vicinity of $\msq3l$. However, as mentioned earlier,
since the mixing in the stop sector is driven by $\mtop$, $A_t$ plays
a crucial role in mixing and help achieve close to maximal mixing 
even when $\mstopr$ is somewhat away from $\msq3l$. This is clear from
the bluish-purple and yellowish-orange bands extended over the range
700 GeV $< \mstopr <$850 GeV.
\item For smaller values of $\mstopr$ ($\gtrsim 450$ GeV), $\mstop1$ required by
us is close to these values. Hence, for this range, $\sstop1$ is almost purely
right-chiral ($\thetastop \simeq 90^\circ$). No mixing is required for
the purpose and hence $A_t$ values are seen to be within a couple of 
hundred GeVs.
\end{itemize}

The problem is now to understand if the imprints of these branching
fractions can be read out by studying suitable final states at the LHC. 
As indicated earlier, this is going to be rather challenging given that
the third generation squarks all lead to very similar final states.
Thus, disentangling individual contributions (read, contaminations), 
which is so necessary to unravel the sector, is a complicated  
proposition. This is more so since, in a bottom-rich environment, 
identifying multiple bottom quarks could very well hold the key.
%

To simulate SUSY events we use the event generator Pythia v6.420 
\cite{Sjostrand:2006za}. 
Pythia is interfaced with the framework SUSYHIT \cite{Djouadi:2006bz}
 that in turn uses 
the SLHA \cite{Allanach:2008qq} protocol to integrate Suspect 2.31 
\cite{Djouadi:2002ze}, 
the popular SUSY mass-spectrum generator 
and SDECAY \cite{Muhlleitner:2003vg} 
and HDECAY \cite{Djouadi:1997yw}
which calculate the decay branching fractions of the SUSY 
particles and the Higgs bosons respectively. 
To simulate the SM backgrounds discussed in 
section 3, partonic events are generated with Alpgen v2.13 \cite{Mangano:2002ea} 
We have used CTEQ6L \cite{Pumplin:2002vw} parametrisation of the parton
distribution function (PDF) via Pythia's interface to LHAPDF v5.7 
\cite{Whalley:2005nh}.
The renormalisation/factorisation scale is set to $\sqrt{\hat{s}}$,
for both signal and background analyses.
In case of background processes in Alpgen for which $\sqrt{\hat{s}}$ 
is not available as a choice for the renormalisation/factorisation scale, 
the default option for the same is used.
To attempt a somewhat realistic treatment of the final state objects
like jets, leptons, photons and the missing transverse energy
we interfaced AcerDET v1.0 \cite{RichterWas:2002ch} as the fast 
detector simulator. 

Unweighted events for the signal processes from Pythia and that for the 
SM background processes from Alpgen are then showered using Pythia with 
initial and final state radiations on. To avoid possible double-counting, 
multijet events from the Matrix Element (ME) calculation in Alpgen are 
matched with the jets from the Parton Shower (PS) using the MLM 
prescription \cite{Hoche:2006ph}, available in Alpgen as the default.
In the entire exercise, a jet is defined with a minimum clustered 
energy in the calorimeter of 25 GeV having a cone size of 
$\Delta R=0.4$  within $|\eta| \leq 2.5$.

For a jet to be triggered as a $b$-jet, it is required to have $p_T>5$ 
GeV within  $|\eta|<2.5$ and an isolation from a neighbouring jet of
$\Delta R < 0.2$ is required. Further, an average tagging
efficiency of 50\% is used for the $b$-jets. 

Electrons and muons are triggered in AcerDET only if they have 
$p_T>5$ GeV, $|\eta|<2.4$ and have a lepton-jet separation
$\Delta R_{\ell j} > 0.4$. Further, to ensure the purity of the
candidates, energy deposited within a cone of radius 
$\Delta R < 0.2$ about the candidate lepton was 
required to be within 10 GeV.

We incorporate a generic set of kinematic cuts in our analysis following ATLAS 
\cite{Aad:2009wy}. The requirements for the final state objects and the employed 
cuts are as follows:
\begin{itemize}
\item Two tagged $b$-jets (with a $b$-tagging efficiency of 50\%), 
      both with $p_T > 40$ GeV.
\item At least four jets with the hardest one requiring $p_T^j > 100$ GeV 
      and the other three with $p_T^j> 50$ GeV. On top of this, for an inclusive 
      jet final state all jets should have $p_T^{jet} > 40$ GeV.
\item For inclusive 1-lepton and SSDL final states, isolated leptons with 
      $p_T^\ell > 20$ GeV are required. For OSDL final state a, both leptons
      are required to have $p_T^\ell > 10$ GeV. Also, for the SSDL and OSDL
      final states two and only two leptons are required.
\item missing $\etmiss >150$ GeV,
\end{itemize}

On top the above set of cuts, we use cuts on two more kinematic variables which
help reduce the SM background in an effective way. The first one is
the so-called \emph{transverse mass} of the system comprised of the lepton(s) 
and missing energy and defined as 
\[
M_{_{T}} = \sqrt{(E^\ell_{T} + {\not \! \! E}_T)^2  
         -  ( p^\ell _{x} + {\not \! p}_x )^2  
         -  ( p^\ell _{y} + {\not \! p}_y )^2}. 
\]
Traditionally used to reconstruct a leptonically decaying $W$-boson, a suitable
cut on $M_T$ thus could efficiently reduce the $W$-boson background from the
SM processes. In the left panel of Figure \ref{figure:mT-meff} we show the
$M_T$ distributions for the two benchmark scenarios of Tables 
\ref{table:high-mass} and \ref{table:low-mass} and that for the SM background 
combined over all the contributing processes. The $m_T$ profile for the SM 
background (the profile in light-green) nose-dives beyond 100 GeV with
a subdominant tail extending up to 500 GeV. For the
signal, the one for the high-mass case (in yellow) and the low-mass one (in blue) 
extend to 550 GeV and 650 GeV, respectively. As we can see, a cut of 
$M_T > 200$ GeV would efficiently eliminate the SM background where $W$ bosons 
are associated.
%
\begin{figure}[htbp]
\centering
\mbox{
\hspace*{-1.0cm} \includegraphics[height=8.5cm,width=12.5cm,angle=0]{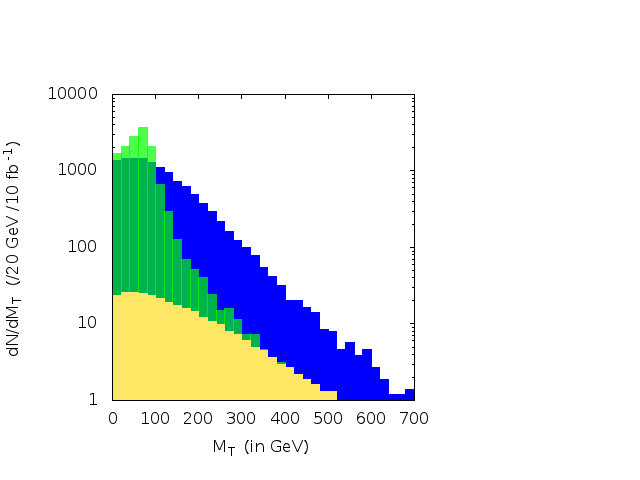}
\hspace*{-2.8cm} \vspace*{-2.0cm} \includegraphics[height=7.0cm,width=7.4cm,angle=0]{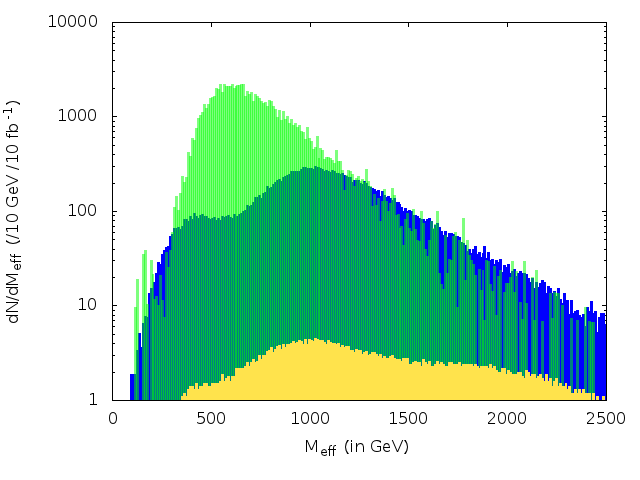}}
\caption{\small 
The transverse mass (left) and the effective mass (right) spectra for the summed-up 
background (in light green/light grey) and the signal. The profile at the bottom in 
yellow/ash is for the signal in the high-mass benchmark scenario of 
Table \ref{table:high-mass}.
The profile in blue/black is for the signal in the low-mass benchmark scenario of 
Table \ref{table:low-mass}.}
\label{figure:mT-meff}
\end{figure}

It is also known that the variable called effective mass ($M_{eff}$) where
\[ M_{eff} = \sum_{\!\!\!\!\!\!\!\!{_{_{_{jets}}}}} p_T^j 
 + \sum_{\!\!\!\!\!\!\!\!\!\!{_{_{_{\ell eptons}}}}} p_T^\ell + \etmiss \]
can also be a powerful generic discriminator in search of new physics with
highly massive exotic states. In the right panel of Figure \ref{figure:mT-meff} 
we illustrate the $M_{eff}$ distributions for the signal and the SM background 
as described for the $M_T$ curves and following the same colour convention. 
Studying the trends of the distributions in this plot, we require a flat minimum of 
800 GeV for $M_{eff}$ for analysing both high and low mass benchmark scenarios. 
In the latter case, it seems to be a little overkill though.
However, we stick to this particular value to bring uniformity in the overall
analysis (the SM background thus remaining the same) which indeed may render 
this part of the study only conservative.

In the subsequent part of this section we demonstrate how the event counts
for different final states vary not only as a function of $\thetastop$ but also
when $\thetasbot$ is varied while $\msbot1$ and $\mstop1$ are held fixed within
$\pm 10$ GeV of the corresponding benchmark values. In all these cases we
incorporated the kinematic cuts and the the $b$-tagging efficiency as mentioned
above.

The variations in the mixing angles are effected by simultaneously tweaking 
$\msq3l$, $\mstopr$, $\msbotr$ and $A_t$. The change in rates in 
different multi-lepton and jet final states reflect the sensitivity of the 
rates to varying mixing angles, masses being held fixed. For this, we again 
take the spectra of Tables \ref{table:high-mass} and \ref{table:low-mass}
as the reference ones with somewhat higher and lower squark masses, respectively.

In Table \ref{table:high-mass-uni} we present the different event rates for the
high-mass scenario of Table \ref{table:high-mass}. Note that the benchmark scenario 
here is chosen in a way such that 
$\thetastop\approx 67^\circ$ which means the mixing angle is halfway between 
the maximal (45$^\circ$) and what corresponds to a purely right-handed 
$\sstop1$ (90$^\circ$). To be precise, $\thetastop=67^\circ$ refers to about 
a 15\% admixture of $\stopl$ in $\sstop1$. The mixing in the sbottom sector 
remains practically fixed at a rather low angle ($\approx 2^\circ$) thus 
rendering $\sbot1$ to be almost purely left-chiral. 

%
\begin{table}[htb]
\begin{center}
\begin{tabular}{||c|c||c|c|c||}
\hline
\hline
$\thetastop , \; \thetasbot$ & 
BR[$\sbot1 \to \sstop1 W^-$] &
2$b$ + $\geq 4j$ +$\etmiss$ & 
2$b$ + $\geq 4j$ + $\etmiss$  & 
2$b$ + $\geq 4j$ + $\etmiss$ \\
& & $\geq 1 \ell$  & + $OSDL$ & + $SSDL$ \\\hline
\hline
25.1$^\circ$ ,  4.7$^\circ$ & 89.4 \% & 66  & 11 & 4 \\
$\sstop1^{L_{max}}$ , $\sbot1^L$  & & & & \\
\hline
46.6$^\circ$ ,  46.0$^\circ$  & 78.8 \% & 74 & 15 & 5 \\
$\sstop1^{LR}$ , $\sbot1^{LR}$  & &  &  &  \\
\hline
33.5$^\circ$ ,  83.0$^\circ$ & 45.5 \% & 48 & 7 & 2 \\
$\sstop1^{L_{max}}$ , $\sbot1^R$  & &  &  &  \\
\hline
From Table 1 &  &  &  & \\
67.0$^\circ$ ,  1.9$^\circ$ & 64.0 \% & 65 & 14 & 4 \\
$\sstop1^{L{\boldsymbol R}}$ , $\sbot1^L$  & &  &  &  \\
\hline
84.7$^\circ$ ,  2.6$^\circ$ & 9.4 \% & 58 & 9 & 3 \\
$\sstop1^R$ , $\sbot1^L$  & &  &  &  \\
\hline
87.9$^\circ$ ,  87.2$^\circ$ & 0.0 \% & 27 & 3 & 0 \\
$\sstop1^R$ , $\sbot1^R$  & &  &  &  \\
\hline
\hline
SM Background & - & 12 & 4 & 0 \\
\hline
\hline
\end{tabular}
\end{center}
\caption{Variations of events rates (for the spectrum in Table \ref{table:high-mass})
in different leptonic final states
containing two tagged $b$-jets along with other light quark jets and
missing transverse energy with different combinations of $\thetastop$
and $\thetasbot$. The kinematic cuts used are as discussed in the text. 
The event-rates correspond to an accumulated luminosity of 300 fb$^{-1}$
and an average $b$-tagging efficiency of 50\%.
The last row presents the summed-up events rates from various SM 
backgrounds (see text for details) for the respective final states
under the same set of cuts.}
\label{table:high-mass-uni}
\end{table}
%
%
In the first column of Table \ref{table:high-mass-uni}, he mixing angles in 
the stop and the sbottom sectors are presented along with the corresponding 
chiral contents of $\sstop1$ and $\sbot1$. The suffix $LR$ on $\sstop1$ and 
$\sbot1$ indicates that these mass eigenstates have some significant $L$ 
and $R$ contaminations while either $L$ or $R$ as a suffix indicates that 
these are either almost purely left- or right-chiral in nature, respectively. 
It is important to note the mixing angles in the first column in this Table. 
These combinations of angles are again obtained by varying 
$m_{\widetilde{Q}_{3L}}$, $\msbotr$, $\mstopr$, $A_t$ and $A_b$ while holding 
$\mstop1$ and $\msbot1$ within $\pm 10$ GeV of the respective benchmark 
values. Same procedure is taken for subsequent Tables in this section.
Although our intention has been to demonstrate the effect of different 
combinations of mixing angles in the sbottom and the stop sectors spanning 
over purely left-chiral pairs to right-chiral ones passing through 
intermediate situations, we cannot really realise a situation when both 
$\sbot1$ and $\sstop1$ are almost pure left-chiral states and still allowing 
for the necessary splitting between $\msbot1$ and $\mstop1$. The reason is 
quite clear and is as follows. Since both $\sbotl$ and $\stopl$ are states 
determined by $\widetilde{Q}_{3_{L}}$, the mass-eigenstates $\sbot1$ and 
$\sstop1$ dominated respectively by them are bound to be rather degenerate 
($\sim m_{\widetilde{Q}_{3_{L}}}$). The only  splitting between them is due 
to the $SU(2)$ $D$-term ($\propto m_Z^2 \cos 2\beta$) originating from the 
different electroweak (isospin ($T_3$), electric charge) quantum numbers the 
left-chiral states carry. The splitting between $\msbot1$ and $\mstop1$ 
(i.e., the one that ensures $\msbot1 > \mstop1 + W$) that we are particularly 
interested in, in the present study, is thus not achievable unless we allow 
for some mixing in the top squark sector. This is what is reflected in the 
first row of Table \ref{table:high-mass-uni}. The angle $\thetastop \approx 
24.7^\circ$ indicates that $\sstop1$ has some right-chiral admixture but still 
dominated (since, $\thetastop < 45^\circ$) by the left-chiral state. This 
observation would have ramifications beyond the present context. It implies 
that given a mass-splitting between $\sbot1$ and $\sstop1$, combination of 
arbitrarily small $\thetasbot$ and $\thetastop$ (i.e., arbitrarily large 
\emph{left} chiral components in both $\sbot1$ and $\sstop1$) are not feasible. 
This is true even when we minimally supersymmetrise the SM without
subjecting it to further constraints. This is a kind of entanglement we 
like to point out. In other words, the phenomenon rules out the possibility 
of having a full-strength coupling in the on-shell decay 
$\sbot1 \to \sstop1 W$ (or, for that matter, $\sstop1 \to \sbot1 W$, for a 
reverse hierarchy of masses between $\sstop1$ and $\sbot1$).

From Table \ref{table:high-mass-uni} it can be seen that the variations in
the event counts of different leptonic final states as the combination of
mixing angles vary are anything but drastic. This is not unexpected.
The reason behind this is that when the branching fraction of 
$\sbot1 \to \sstop1 W$ decreases, the same in other decay modes, 
e.g., $\sbot1 \to t \ch1$ and $\sbot1 \to b \ntrl2$, start increasing
(see Figure \ref{figure:theta-br}).
The decay width for $\sbot1 \to \sstop1 W$ 
could be directly affected by double suppression from both $\thetasbot$ and 
$\thetastop$ at the same time while its other modes of decay only see 
$\thetasbot$ directly. Thus, there can be a possible sharing of 
branching fractions among these modes all of which contribute to
the leptonic final states discussed here. Hence, we may reasonably expect
a less drastic variation of the event counts with varying $\thetasbot$ 
and $\thetastop$. when these modes are open.

Thus, an appropriate setup to improve the sensitivity to the mixing angles 
and hence to see the tell-tale signatures of the variations 
of mixing angles on the event rates of various multilepton final states is to
have a situation when the decay modes $\sbot1 \to t \ch1$ and 
$\sbot1 \to b \ntrl2$ are kinematically closed. This we realise by making
$M_2$ larger than $\msbot1$ ($M_2 = 750$ GeV) so that the lighter chargino
and the second lightest neutralino becomes heavier than $\sbot1$. We also set
$\mu=1$ TeV (changing it from 750 GeV, as was for the benchmark scenario of
Table \ref{table:high-mass-uni}) so that the compositions of the chargino
and the neutralino sectors remain more or less unchanged. 

The resulting variations are presented in Table \ref{table:high-mass-nonuni}.
Branching fractions are now only shared between the decay modes
$\sbot1 \to \sstop1 W$ and $\sbot1 \to b \ntrl1$. 
Events from these two decay modes could, to a good extent, may unambiguously
tell us about the respective decay branching fractions which, in turn,
would be indicative of the mixing angles involved. However, 
$\sbot1 \to b \ntrl1$ would lead to leptonically quiet events with
$b$-jets and $\etmiss$ which are hard to identify over the huge SM background,
particularly damaging one being of QCD origin with `fake' $b$-jets.
Thus, one needs to concentrate on how the rates in different multilepton
final states are varying to get an idea about the mixing angle(s) in the 
sbottom and stop sectors. 
The combinations of angles
are kept close to the corresponding ones of Table \ref{table:high-mass-uni}.
As we can see, the branching fractions to $\sstop1 W$ mode, shown in column 2, 
displays a much clearer variation with the angle-combinations when compared to
Table \ref{table:high-mass-uni}.
With increasing right-chiral component in both $\sbot1$ and $\sstop1$,
the branching fraction $\sbot1 \to \sstop1 W$ gets diminished straight-away.
Note that this decay mode of sbottom is the only source of leptons from $\sbot1$
down the cascade. The other one being $\sbot1 \to b \ntrl1$, this would
only lead to $b$-jets and missing energy in the final state. Thus, a more drastic
(compared to Table \ref{table:high-mass-uni} decrease in the number of events in 
different leptonic final states is expected as right chiral components increase. 
This is exactly what Table \ref{table:high-mass-nonuni} indicates.
%

\begin{table}[htb]
\begin{center}
\begin{tabular}{||c|c||c|c|c||}
\hline
\hline
$\thetastop \;, \; \thetasbot$ & 
BR[$\sbot1 \to \sstop1 W^-$] &
2$b$ + $\geq 4j$ +$\etmiss$ & 
2$b$ + $\geq 4j$ + $\etmiss$  & 
2$b$ + $\geq 4j$ + $\etmiss$ \\
& & + $\geq 1 \ell$   & + $OSDL$ & + $SSDL$ \\
\hline
\hline
24.7$^\circ$ ,  2.1$^\circ$ & 99.7 \% & 47  & 9 & 4 \\
$\sstop1^{L_{max}}$ , $\sbot1^L$ &  & & & \\
\hline
45.6$^\circ$ ,  44.8$^\circ$ & 97.1 \% & 49  & 10 & 4 \\
$\sstop1^{LR}$ , $\sbot1^{LR}$ & &  & & \\
\hline
39.4$^\circ$ ,  87.2$^\circ$ & 9.9 \% & 10 & 1 & 0 \\
$\sstop1^{L_{max}}$ , $\sbot1^R$  & & & & \\
\hline
88.5$^\circ$ ,  2.8$^\circ$ & 27.8 \% & 16 & 3 & 0 \\
$\sstop1^R$ , $\sbot1^L$  & & & & \\
\hline
89.2$^\circ$ ,  87.5$^\circ$ & 0.0 \% & 7 & 1 & 0 \\
$\sstop1^R$ , $\sbot1^R$  & & & & \\
\hline
\hline
SM Background & - & 12 & 4 & 0 \\
\hline
\hline
\end{tabular}
\end{center}
\caption{The same variations as shown in Table \ref{table:high-mass-uni} 
but with nonuniversal gaugino masses where for $\sbot1$, only the modes 
$\sbot1 \to \sstop1 W$ and $b \ntrl1$ are open. Except for $M_2=750$ GeV 
and $\mu=1$ TeV all other parameters are as there in 
Table \ref{table:high-mass}.}
\label{table:high-mass-nonuni}
\end{table}
%

In both Tables \ref{table:high-mass-uni} and \ref{table:high-mass-nonuni},
the rates for both OSDL and SSDL final states are found to be on the 
lower side compared to the inclusive 1-lepton case. This is quite expected 
since final states with more number of leptons involve further suppression due
to added leptonic branching. The combined SM backgrounds for the respective 
final states are indicated in the last rows of both the Tables. As can be 
gleaned from these numbers, except for the cases where both $\sbot1$ and 
$\sstop1$ have significant right-chiral components, the rates in the 
inclusive 1-lepton final state has significance above 5$\sigma$ at an 
accumulated luminosity of 300 fb$^{-1}$. 
On the other hand, only the OSDL rate with both $\sstop1$ and $\sbot1$ 
having maximal possible left-chiral contamination (i.e., $\sstop1^{LR}$, 
$\sbot1^L$ combinations in the first row) passes 
the 5$\sigma$ mark\footnote{This assumes a 
Gaussian estimation of the significance. However, for some of these
low yields a Poissonian treatment would be more appropriate. The main issue here, 
however, is to take a simple note of the depleting counts in the dilepton 
final states for an integrated luminosity of 300 fb$^{-1}$.}
at 300 fb$^{-1}$ 
The signal rates in the SSDL mode are indeed low throughout. However,
since the SM background for this can be virtually eliminated, seeing a 
few events would suffice. Note that, at this stage, even a two fold increase 
in the accumulated luminosity is not going to change the overall situation
drastically, at least qualitatively, except for the fact that we may 
have then a handful of very clean SSDL events. The bottom-line of Table 
\ref{table:high-mass-nonuni} is that 
there could be a clear imprint of the product of $\cos\thetasbot$ and
$\cos\thetastop$ in the absolute and mutually relative rates of these
leptonic final states, though at a somewhat high integrated luminosity. 
Thus, the corresponding mass value of the sbottom squark (around 700 GeV,
along with the gluino of 1200 GeV) lives on the edge of explorability for 
such a study. This prompts us to take up an exactly similar study but now
with a low-lying spectrum benchmarked in Table \ref{table:low-mass} and
its non-universal (in terms of gaugino mass relationship) variant. 

In Tables \ref{table:low-mass-uni} and \ref{table:low-mass-nonuni} we present 
studies which exactly emulate the proceedings of Tables \ref{table:high-mass-uni} and
\ref{table:high-mass-nonuni}, respectively but using the spectrum of Table 
\ref{table:low-mass}. 
With $\msbot1 \approx 500$ GeV and $\mgluino \approx 650$ GeV, we definitely
expect a larger yield of events in any given final state. That the numbers
presented in both Tables \ref{table:low-mass-uni} and \ref{table:low-mass-nonuni} 
are of the same order as for their high-mass counterparts of Tables
\ref{table:high-mass-uni} and \ref{table:high-mass-nonuni}, reflects the fact that
the former Tables have numbers for a much lower integrated luminosity of only 
30 fb$^{-1}$ as compared to 300 fb$^{-1}$ for the latter ones. SM backgrounds
also get scaled down by this luminosity factor. Tables
\ref{table:low-mass-uni} and \ref{table:low-mass-nonuni} clearly demonstrate how
the sensitivities to mixing angles can be better studied with lower masses of
the involved SUSY particles for all the final states discussed.

%
\begin{table}[htb]
\begin{center}
\begin{tabular}{||c|c||c|c|c||}
\hline
\hline
$\thetastop , \; \thetasbot$ & 
BR[$\sbot1 \to \sstop1 W^-$] &
2$b$ + $\geq 4j$ +$\etmiss$ & 
2$b$ + $\geq 4j$ + $\etmiss$  & 
2$b$ + $\geq 4j$ + $\etmiss$ \\
& & $\geq 1 \ell$  & + $OSDL$ & + $SSDL$ \\\hline
\hline
31.4$^\circ$ ,  2.7$^\circ$ & 57.1 \% & 36  & 12 & 4 \\
$\sstop1^{L_{max}}$ , $\sbot1^L$  & & & & \\
\hline
45.7$^\circ$ ,  47.7$^\circ$  & 45.9 \% & 42 & 11 & 5 \\
$\sstop1^{LR}$ , $\sbot1^{LR}$  & &  &  &  \\
\hline
32.3$^\circ$ ,  82.7$^\circ$ & 16.6 \% & 32 &  8 &  1 \\
$\sstop1^{L_{max}}$ , $\sbot1^R$  & &  &  &  \\
\hline
From Table 2  & & & &  \\
64.8$^\circ$ ,  2.7$^\circ$ & 29.4 \% &  46 &  11 &  5 \\
$\sstop1^{L{\boldsymbol R}}$ , $\sbot1^L$  & &  &  &  \\
\hline
86.6$^\circ$ ,  3.2$^\circ$ & 1.1 \% & 58 & 16 & 7 \\
$\sstop1^R$ , $\sbot1^L$  & &  &  &  \\
\hline
79.7$^\circ$ ,  88.8$^\circ$ & 0.0 \% &  38 &  10 & 3 \\
$\sstop1^R$ , $\sbot1^R$  & &  &  &  \\
\hline
\hline
SM Background & - &  1 & 0 & 0 \\
\hline
\hline
\end{tabular}
\end{center}
\caption{Variations of events rates (for the spectrum in Table \ref{table:low-mass})
in different leptonic final states
containing two tagged $b$-jets along with other light quark jets and
missing transverse energy with different combinations of $\thetastop$
and $\thetasbot$. The kinematic cuts used are as discussed in the text. 
The event-rates correspond to an accumulated luminosity of 30 fb$^{-1}$
and an average $b$-tagging efficiency of 50\%.
The last row presents the summed-up events rates from various SM 
backgrounds (see text for details) for the respective final states
under the same set of cuts.}
\label{table:low-mass-uni}
\end{table}
%
%
\begin{table}[htb]
\begin{center}
\begin{tabular}{||c|c||c|c|c||}
\hline
\hline
$\thetastop \;, \; \thetasbot$ & 
BR[$\sbot1 \to \sstop1 W^-$] &
2$b$ + $\geq 4j$ +$\etmiss$ & 
2$b$ + $\geq 4j$ + $\etmiss$  & 
2$b$ + $\geq 4j$ + $\etmiss$ \\
 & & + $\geq 1 \ell$   & + $OSDL$ & + $SSDL$ \\
\hline
\hline
34.9$^\circ$ ,  2.5$^\circ$ & 97.8 \% &75  & 24 & 5 \\
$\sstop1^{L_{max}}$ , $\sbot1^L$  & & & & \\
\hline
47.8$^\circ$ ,  47.3$^\circ$ & 86.3 \% & 69  & 19 & 6 \\
$\sstop1^{LR}$ , $\sbot1^{LR}$ & & & & \\
\hline
32.3$^\circ$ ,  82.6$^\circ$ & 20.1 \% & 51 & 12 & 5 \\
$\sstop1^{L_{max}}$ , $\sbot1^R$  & & & & \\
\hline
60.7$^\circ$ ,  2.9$^\circ$ & 94.9 \% & 73 & 22 & 7 \\
$\sstop1^{L{\boldsymbol R}}$ , $\sbot1^L$  & & & & \\
\hline
81.8$^\circ$ ,  2.5$^\circ$ & 60.6 \% & 70 & 15 & 7 \\
$\sstop1^R$ , $\sbot1^L$  & & & & \\
\hline
80.8$^\circ$ ,  88.8$^\circ$ & 0.0 \% & 44 & 9 & 3 \\
$\sstop1^R$ , $\sbot1^R$  & & & & \\
\hline
\hline
SM Background & - & 1 & 0 & 0 \\
\hline
\hline
\end{tabular}
\end{center}
\caption{The same variations as shown in Table 5 but with nonuniversal 
gaugino masses where for $\sbot1$
decay, only the modes $\sbot1 \to \sstop1 W$ and $b \ntrl1$ are open.
Except for $M_2=550$ GeV and $\mu=700$ GeV, values of all other parameters 
are same as in Table \ref{table:low-mass}.}
\label{table:low-mass-nonuni}
\end{table}
%

The results presented in Tables \ref{table:low-mass-uni} and \ref{table:low-mass-nonuni}
have direct correspondences to their high-mass counterparts presented in Tables 
\ref{table:high-mass-uni} and \ref{table:high-mass-nonuni}.
Note that, in Tables \ref{table:high-mass-nonuni} and \ref{table:low-mass-nonuni},
even when both $\sstop1$ and $\sbot1$ are both almost purely right chiral, there 
are still significant number of leptonic (in particular, inclusive one-lepton 
events) events present, which is not quite 
expected had $\sbot1 \to \sstop1 W$ been the sole source of leptons in the final 
state. In fact, in our simulation not only $\sbot1$-pairs but also 
$\gluino\gluino$ and $\gluino \sbot1$ pairs are included, as pointed out in the 
beginning of section 3. Thus, most of the leptons, under such a circumstance, are 
coming from top-squarks produced in the decay of the gluino along with the ones 
from the decay of SM top quark obtained under the SUSY cascade. 
This brings us to an important issue. 

Of particular interest is the cascade 
$\sbot1 \to \sstop1 \, W^-\to t \,\ntrl1 \; W^-\to b \, W^+\,\ntrl1\;W^-
\to b \, + \; \ell eptons \: and/or \: jets \: + \, \etmiss$.
Here, note that the decay of a single $\sbot1$ gives rise to two $W$-bosons
(with opposite charges): one coming directly from the decay of $\sbot1$
while the other appearing in the decay of a top quark (originating from
a top squark) further down the cascade. Thus, if both $\sbot1$-s undergo 
such a cascade, at some stage there would be four (4) $W$  bosons there. 
Possibility of identifying more of them could provide us with a remarkable 
handle to remove the above-mentioned `spurious' leptons that do not originate 
in the cascades of $\sbot1$. Even the 
model-background originating from direct (in pairs) production of $\sstop1$,
which has not been considered in this work, can be removed if multiple $W$-s
can be reconstructed in the final states discussed here.
This would facilitate direct probe to the coupling $\sbot1 \sstop1 W$ thus
reflecting on the mixings in the bottom and the top squark sectors.

To the best of our knowledge dedicated study in this line is still lacking. 
One can take useful cue from some recent studies on multi-top final states 
\cite{Acharya:2009gb,Brooijmans:2010tn,Cacciapaglia:2011kz}.
As pointed out in these works, reconstructing multiple top quarks is an 
inherently difficult proposition, particularly in the complex environment of 
the LHC where the final state objects overlap.
However, identifying  multiple $W$-bosons
is not expected to be more complicated than tracking down multiple top
quarks. This is because, identifying multiple top quarks involves successive
reconstructions of first, $W$ bosons and subsequently, the individual
top quarks they are coming from. Also, note that with our choice of soft 
SUSY parameters, $\sstop1$ always
decays to $t \ntrl1$ thus making one of the $W$ bosons in the cascade always
coming from a top quark. This may offer some degree of simplicity while 
undertaking a feasibility-study of exploring such a cascade. It can be 
foreseen that capability of identifying 
these $W$ bosons would prove to be crucial in a generic scenario where
$\sbot1$ can decay to other channels ($t \, \ch1$, $b \ntrl2$, for example)
which ultimately give rise to identical final states.
%

\section{Summary and Outlook}
Physics analyses of recent LHC data hint essentially to super-TeV squarks 
from the first two generations and to a somewhat massive gluino. However, till 
now, the LHC experiments are not as sensitive to searches for the sbottom 
and the stop squarks. Consequently, much lighter sbottom and stop squarks are
still allowed. This is exactly under such a situation when searches 
for them assume special significance. 

%

In this work, we aim for a rather conservative approach. We presume a 
scenario where only the lightest of the bottom and top squarks have 
significant cross sections at the 14 TeV run of the LHC. 
Gluinos are taken to be intermediate in mass such that they
could contribute to our final states only in a moderate way.
On the other hand, squarks from the first two generations 
are considered to be heavier than a TeV following recent analyses.
These are, however, still within the reach of LHC with reasonable
cross sections and are able
to contribute to the final states considered by us in different ways.
However, in the spirit of the present work, to be conservative, we
neglected those contributions. 
The compulsion is then to learn from the limited, nonetheless 
crucial, piece of information offered through the squarks from the 
third generation.

It is pointed out that requiring some splitting between the
the lighter sbottom and the stop eigenstates may set in some kind of
an entanglement between the two sectors. This is because of the common
left-chiral soft mass that enters the diagonal terms of the mass-squared 
matrices for both the sectors.
One possible fall out of such entanglement is that the
mixing angle in one of these sectors may constrain the corresponding 
one in the other sector. The degree of such entanglement may depend
upon quite a few factors. The absolute masses of $\sbot1$ and $\sstop1$
and the splitting between these masses are two such important ones.

In this work we demonstrate the phenomenology of the lighter
sbottom squark. The reference decay mode considered is $\sbot1 \to \sstop1 W$.
This decay mode is naturally enhanced when both $\sbot1$ and $\sstop1$
have significant left-chiral content in them. Sensitivity of the resulting 
phenomenology to variations of the chiral contents of 
$\sbot1$ and $\sstop1$ would thus help probe their compositions.
We point out the situations under which a clean study is possible.
Under involved situations, we suggest that our ability to identify more number
of $W$-bosons in the final state would hold the key. We also stress that
observations in multiple final states naturally would facilitate understanding.
We demonstrate the issue with somewhat heavy sbottom and stop followed by the 
lighter ones at the 14 TeV LHC run. For the heavier spectrum we required
an integrated luminosity in the ballpark of 300 fb$^{-1}$ while for the lighter
spectrum 30 fb$^{-1}$ may be good enough.

Maximally (even moderately) mixed top and bottom squarks is something outside
the realm of much popular CMSSM/mSUGRA frameworks. Any hint of such mixings
would firmly indicate departure from these scenarios.  As exemplified in 
Tables \ref{table:high-mass-nonuni} and \ref{table:low-mass-nonuni}, 
it may happen, that such a study may crucially bank on the nonuniversal 
masses in the gaugino sector thus exposing another crucial piece of 
information on the SUSY spectrum in the same go.

A further entanglement is envisaged between the chargino/neutralino sector 
and the sector comprising of the third generation squarks. Here, the common 
agents responsible are $\mu$ and $\tan\beta$. The role of $\mu$ in such 
entanglements is expected to be tempered by the value of $\tan\beta$ and 
hence could well be limited to scenarios with light stop and sbottom. 
Further, due to the opposite roles played by $\tan\beta$ in determining the 
mixings in the stop and sbottom sectors, sbottom sector may see somewhat 
larger correlation with the chargino/neutralino sector. It is important
to note that since both mixings and masses in the latter sector are 
controlled (to different extents) by $\mu$ and $\tan\beta$, the entanglements 
can easily have both kinematic and dynamical implications. Moreover, with 
squarks from the third generations involved, they respond to both gaugino 
and higgsino components of the charginos/neutralinos in characteristic ways 
under SUSY cascades.

In this work, we have not considered a substantial, positive higher order 
correction (at NLO or NLO+NLL combined) to the SUSY cross sections.
Consideration of this, could increase the cross section significantly 
(by $\sim 35\%$) for our benchmark scenarios. However, to be conservative 
in our approach, and that we are unable to consistently treat the SM 
background on a similar footing, we postpone this to a future work.
Further, we have not included possible contributions from $\sbot2$ and $\sstop2$
either from direct production or from cascades. The former can very well be
small while contribution from the cascades may still be appreciable under 
favourable circumstances. Neither do we  discuss a possible reverse hierarchy 
of $\mstop1 > \msbot1$ where a similar phenomenology with stress on the decay 
$\sstop1 \to \sbot1 W$ can be studied. Also, a dedicated analysis could have been 
undertaken for the 7-TeV run of the LHC. Apriori, given that the cross sections 
for the processes relevant for such a study are much smaller at 7 TeV, only some 
legitimate low-lying spectra might be of interest. However, even in that case,
the requirement for total integrated luminosity would be in the ballpark of 
$\lesssim 100 \; {\mathrm{fb}}^{-1}$ which the LHC is not foreseeing for its 
7-TeV run.

\vskip 15pt
\noindent
{\bf Acknowledgements:}
The authors are partially supported by funding available from 
the Department of Atomic Energy, Government of India for the 
Regional Centre for Accelerator-based Particle Physics (RECAPP), 
Harish-Chandra Research Institute. AD likes to thank J-L Kneur, 
M. Muhlleitner and D. Zerwas for help in using the packages 
SUSPECT and SUSYHIT. Saurabh Niyogi likes to thank Sanjoy Biswas, 
Nishita Desai, Satyanarayan Mukhopadhyay for many useful 
discussions on physics and computational issues. Both AD and SN 
like to thank Amitava Datta for insightful discussions. The authors
acknowledge the use of computational facility available 
at RECAPP and thank Joyanto Mitra for technical help. 
%

%
\end{document}